\documentclass[11pt]{article}
\title{Total Variation Minimization in Compressed Sensing}
\author{Felix Krahmer, Christian Kruschel, and Michael Sandbichler}

\usepackage{amsthm, amssymb, amsmath}
\usepackage{todonotes}
\usepackage{fullpage}
\newcommand{\R}{\mathbb{R}}
\newcommand{\C}{\mathbb{C}}
\newcommand{\Ex}{\mathbb{E}}
\newcommand{\Prob}{\mathbb{P}}
\renewcommand{\S}{\mathbb{S}}

\newcommand{\Ncal}{\mathcal{N}}
\newcommand{\Scal}{\mathcal{S}}
\newcommand{\Ocal}{\mathcal{O}}
\newcommand{\Acal}{\mathcal{A}}
\newcommand{\Bcal}{\mathcal{B}}
\newcommand{\Mcal}{\mathcal{M}}

\newcommand{\Lin}{\mathrm{Lin}}

\newcommand{\identitymatrix}{\mathcal{I}}
\newcommand{\argmin}{\text{argmin}}

\newcommand{\ip}[2]{\langle #1,#2\rangle}

\newcommand{\ch}{\mathrm{ch}}
\newcommand{\diam}{\mathrm{diam}}
\newcommand{\tvsupp}{\mathrm{Jsupp}}
\newcommand{\ee}{\mathrm{e}}

\newcommand{\dx}{\mathrm{d}}

\newcommand{\eps}{\varepsilon}

\newcommand{\Ktv}{K_{TV}^{2\sqrt{s}}}

\theoremstyle{definition}
\newtheorem{theo}{\textsc{Theorem}}[section]
\newtheorem{defi}{\textsc{Definition}}[section]
\newtheorem{lemma}{\textsc{Lemma}}[section]
\newtheorem{coro}{\textsc{Corollary}}[section]

\begin{document}
	\sloppy
\maketitle

\begin{abstract}
This chapter gives an overview over recovery guarantees for total variation minimization in compressed sensing for different measurement scenarios.  In addition to summarizing the results in the area, we illustrate why an approach that is common for synthesis sparse signals fails and different techniques are necessary. Lastly, we discuss a generalizations of recent results for Gaussian measurements to the subgaussian case.
\end{abstract}

\section{Introduction}\label{sec:Introduction}

The central aim of Compressed Sensing (CS) \cite{crt2006robust, do06-2} is the recovery of an unknown vector from very few linear measurements.
Put formally, we would like to recover $x\in\R^n$ from $y = Ax+ e\in \R^m$ with $m\ll n$, where $e$ denotes additive noise. 

For general $x$, recovery is certainly not possible, hence additional structural assumptions are necessary in order to be able to guarantee recovery. A common assumption used in CS is that the signal is  \emph{sparse}. Here for $x$ we assume
\[\Vert x\Vert_0 := |\{k\in[n]\colon\, x_k\neq 0\}|\leq s,\]
that is, there are only very few nonzero entries of $x$. 
And say that $x$ is $s$-sparse for some given sparsity level $s\ll n$.
We call a vector \emph{compressible}, if it can be approximated well by a sparse vector. To quantify the quality of approximation, we let
\[\sigma_s(x)_q:= \inf_{\Vert z\Vert_0 \leq s}\Vert z-x\Vert_q\]
denote the error of the best $s$-sparse approximation of $x$.

In most cases, the vector $x$ is not sparse in the standard basis, but there is a basis $\Psi$, such that $x = \Psi z$ and $z$ is sparse. This is also known as \emph{synthesis sparsity} of $x$. To find an (approximately) synthesis sparse vector, we can instead solve the problem of recovering $z$ from $y = A\Psi z$. 
A common strategy in CS is to solve a basis pursuit program in order to recover the original vector. For a fixed noise level $\eps$, it is given by
\begin{equation}\label{eq:BPDN}
 \text{minimize } \Vert z\Vert_1 \text{ such that } \Vert Az - y\Vert_2\leq \eps.
\end{equation}

\begin{figure}
	\begin{center}
	\includegraphics[width = 0.4\textwidth, height = 0.37\textwidth]{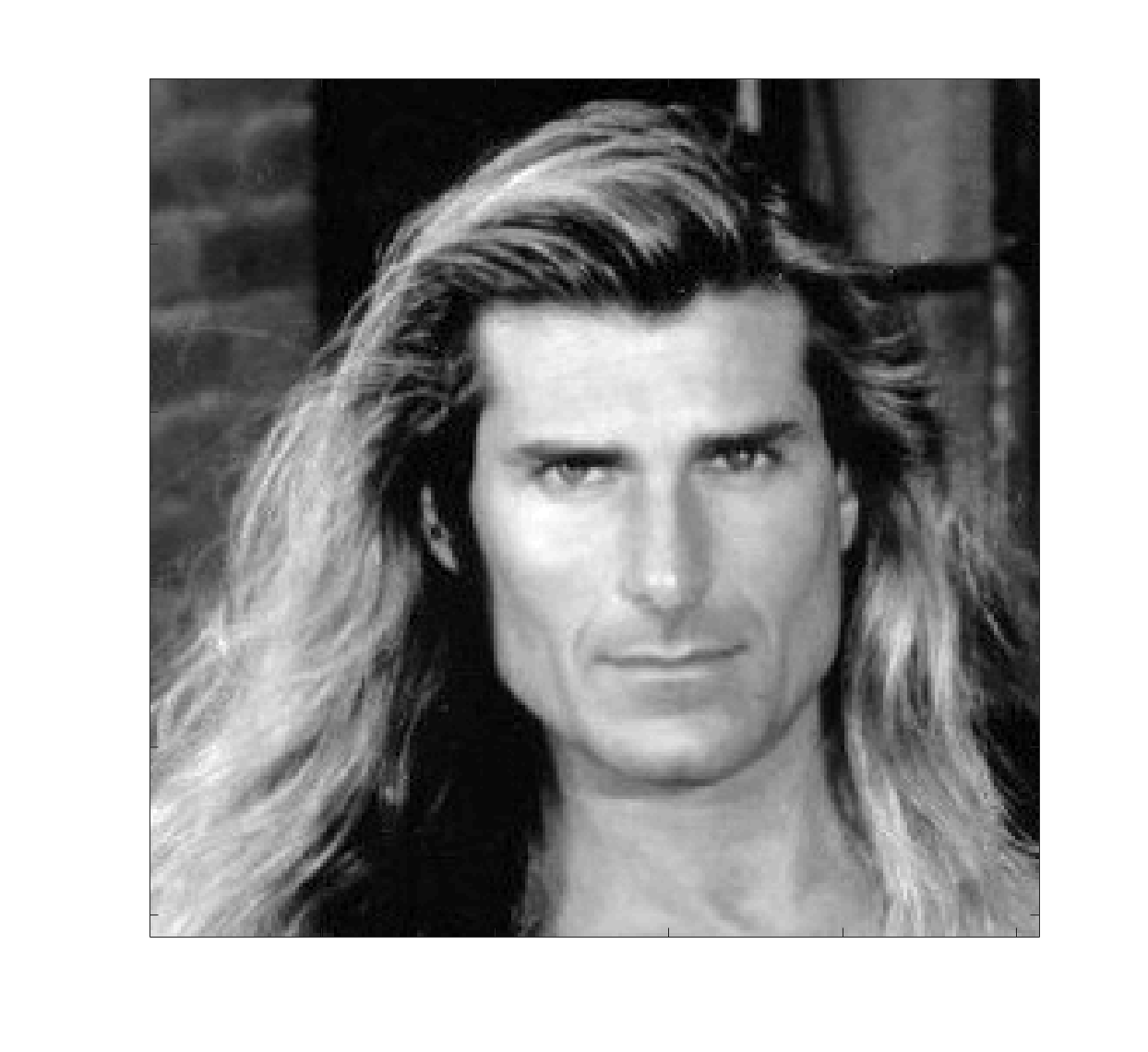}\,
	\includegraphics[width = 0.4\textwidth, height = 0.37\textwidth]{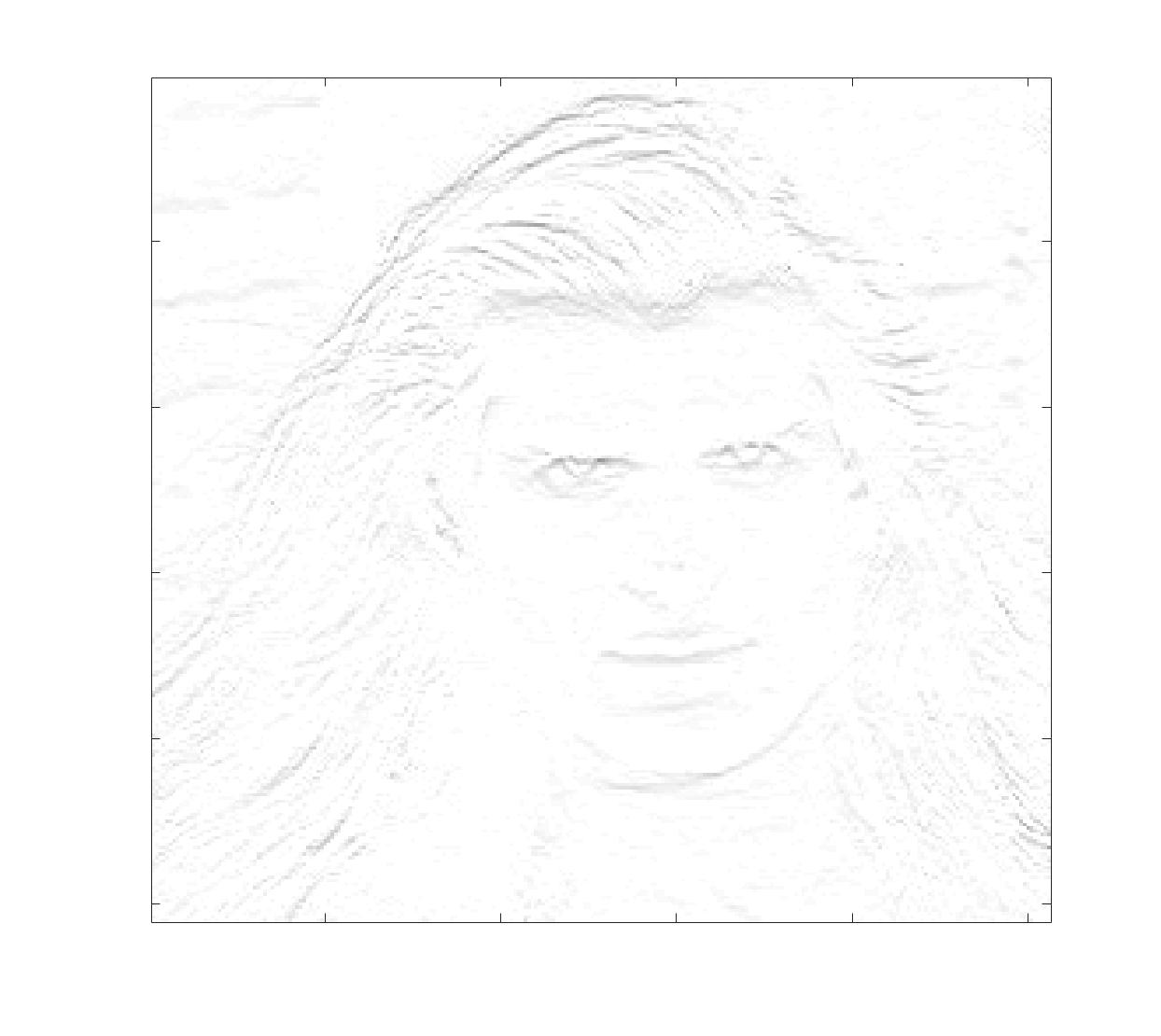}
	\end{center}
	\caption{The original Fabio image (left) and the absolute values after application of a discrete gradient operator(right). }
	\label{fig:DLena}
\end{figure}



While this and related approaches of convex regularization have been studied in the inverse problems and statistics literature long before the field of compressed sensing developed, these works typically assumed the measurement setup was given. The new paradigm arising in the context of compressed sensing was to attempt to use the remaining degrees of freedom of the measurement system to reduce the ill-posedness of the system as much as possible. In many measurement systems, the most powerful known strategies will be based on randomization, i.e., the free parameters are chosen at random.

Given an appropriate amount of randomness (i.e., for various classes of random matrices $A$, including some with structure imposed by underlying applications), one can show that the minimizer $\hat x$ of~\eqref{eq:BPDN} recovers the original vector $x$ with error
\begin{equation}\label{eq:CSerror}
\Vert x - \hat x\Vert_2 \leq c\left(\frac{\sigma_s(x)_1}{\sqrt{s}} + \varepsilon\right),
\end{equation}
see, e.g., \cite{badadewa08} for an elementary proof in the case of subgaussian matrices without structure, and \cite{KR14} for an overview, including many references, of corresponding results for random measurement systems with additional structure imposed by applications.
Note that \eqref{eq:CSerror} entails that if $x$ is $s$-sparse and the measurements are noiseless, the recovery is exact.

For many applications, however, the signal model of sparsity in an orthonormal basis has proven somewhat restrictive. Two main lines of generalization have been proposed. The first line of work, initiated by \cite{rascva08} is the study of sparsity in redundant representation systems, at first under incoherence assumptions on the dictionary. More recently, also systems without such assumptions have been analyzed \cite{CENR10, KNW15}. The main idea of these works is that even when one cannot recover the coefficients correctly due to conditioning problems, one may still hope for a good approximation of the signal.

The second line of work focuses on signals that are sparse after the application of some transform, one speaks of {\em cosparsity} or {\em analysis sparsity} \cite{nam2013cosparse}, see, e.g., \cite{Kabanava2015uniquenessconditions} for an analysis of the Gaussian measurement setup in this framework.
A special case of particular importance, especially for imaging applications, is that of sparse gradients. Namely, as it turns out, natural images often admit very sparse approximations in the gradient domain, see, e.g., Figure~\ref{fig:DLena}. Here the discrete gradient at location $i=(i_1, \dots, i_n)$ is defined as the vector with its $n$ entries given by $\big((\nabla z)_{i}\big)_j = z_{i+e_j} -z_i$, $j=1, \dots, n$, where $e_j$ is the $j$-th standard basis vector.

A first attempt to recover a gradient sparse signal is to formulate a compressed sensing problem in terms of the sparse gradient. When this is possible (for instance in the example of Fourier measurements \cite{crt2006robust}), applying \eqref{eq:BPDN} will correspond to minimizing $\|\nabla z\|_1 =: \|z\|_{TV}$, the {\em total variation seminorm}. Then (under some additional assumptions) compressed sensing recovery guarantees of the form \eqref{eq:CSerror} can apply. This proof strategy, however, only allows for showing that the gradient can be approximately recovered, not the signal. When no noise is present and the gradient is exactly sparse (which is not very realistic), this allows for signal recovery via integrating the gradient, but in case of noisy measurements, this procedure is highly unstable.

Nevertheless, the success motivates to minimize the total variation seminorm if one attempts to recover the signal directly, not the gradient. In analogy with \eqref{eq:BPDN}, this yields the following minimization problem.

\[ \text{minimize } \Vert z\Vert_{TV} = \Vert \nabla z\Vert_1 \text{ such that } \Vert Az - y\Vert_2\leq \eps.\]

For $A$ the identity (i.e., not reducing the dimension), this relates to the famous Rudin-Osher-Fatemi functional, a classical approach for signal and image denoising \cite{rudin1992nonlinear}.
Due to its high relevance for image processing, this special case of analysis sparsity has received a lot of attention recently also in the compressed sensing framework where $A$ is dimension reducing.
%
The purpose of this chapter is to give an overview of recovery results for total variation minimization in this context of compressed sensing (Section~\ref{sec:overview}) and to provide some geometric intuition by discussing the one-dimensional case under Gaussian or subgaussian  measurements (to our knowledge, a generalization to the latter case does not appear yet in the literature) with a focus on the interaction between the high-dimensional geometry and spectral properties of the gradient operator (Section~\ref{sec:RecoverySubgauss}). 

%


\section{An overview over TV recovery results}\label{sec:overview}

In this section, we will give an overview of the state of the art guarantees for the recovery of gradient sparse signals via total variation minimization. We start by discussing in Section~\ref{suffcond} sufficient conditions for the success of TV minimization.

Subsequently, we focus on recovery results for random measurements.
Interestingly, the results in one dimension differ severely from the ones in higher dimensions. Instead of obtaining a required number of measurements roughly on the order of the sparsity level $s$, we need $\sqrt{sn}$ measurements for recovery. 
We will see this already in Subsection~\ref{sec:RecoveryGauss}, where we present the results of Cai and Xu~\cite{cai2015guarantees} for recovery from Gaussian measurements. In Section~\ref{sec:RecoverySubgauss}, we will use their results to obtain refined results for noisy measurements as well as guarantees for subgaussian measurements, combined with an argument of Tropp~\cite{tropp2014convex}.
In Subsection~\ref{sec:RecoveryHaarIncoherent} we will present results by Ward and Needell for dimensions larger or equal than two showing that recovery can be achieved from Haar incoherent measurements.


\subsection{Sufficient Recovery Conditions}\label{suffcond}

Given linear measurements $Ax = y$ for an arbitrary $A\in \R^{m\times n}$ and a signal $x$ with $\|\nabla x\|_0 \le s$,
a natural way to recover $x$ is by solving
\begin{equation}\label{eq:ExactTV}\text{minimize } \| \nabla z\|_1\,\,\text{such that } Az=y.\end{equation}
For $I\subset[n]$ we denote $A_I$ as the columns of $A$ indexed by $I$, and for a consecutive notation we denote $\identitymatrix_{I}^T\nabla$ as the rows of $\nabla$ indexed by $I$ and $\identitymatrix$ as the identity matrix.
The following results can also be easily applied to \emph{analysis $\ell_1$-minimization}, where any arbitrary matrix $D\in\R^{p\times n}$ replaces $\nabla$ in \eqref{eq:ExactTV}, as well as to any real Hilbert space setting \cite{kruschel2015phd}.

In many applications it is important to verify whether there is exactly one solution of~\eqref{eq:ExactTV}.
Since $\nabla$ is not injective here, we cannot easily use the well-known recovery results in compressed sensing \cite{foucart2013mathematical} for the matrix $A\nabla^\dagger$.
However, a necessary conditon can be given since $x$ can only satisfy $Ax = y$ and $(\nabla x)_{I^c} = 0$ if
\begin{align*}
\mbox{ker}(\identitymatrix_{I^c}^T\nabla)\cap\mbox{ker}(A) = \{0\}.
\end{align*}
If $\nabla$ is replaced by the identity, this is equivalent to $A_I$ being injective.
Since this injectivity condition is unavoidable, we assume for the rest of this section that it is satisfied.

The paper~\cite{nam2013cosparse} provides sufficient and necessary conditons for uniform recovery via \eqref{eq:ExactTV}.
The conditions rely on the null space of the measurements and are hard to verify similar to the classical compressed sensing setup~\cite{tillmann2014complexity}.
The following result is a corollary of these conditions. It no longer provides a necessary condition, but is more manageable.
%
%
\begin{coro}\cite{nam2013cosparse}\label{corr:TVNSP}
For all $x\in\R^n$ with $s:= \Vert \nabla x\Vert_0$, the solution of \eqref{eq:ExactTV} with $y=Ax$ is unique and equal to $x$ if for all $I\subset[n]$ with $|I|\le s$ it holds that
\[
\forall w\in\mathrm{ker}(A)\backslash\{0\}\colon\,\,
\|(\nabla w)_I\|_1 < \|(\nabla w)_{I^c}\|_1.
\]
\end{coro}
%
To consider measurements for specific applications, where it is difficult to prove whether uniform recovery is guaranteed, one can empirically examine whether specific elements $x$ solve \eqref{eq:ExactTV} uniquely.
For computed tomography measurements, a \emph{Monte Carlo Experiment} is considered in \cite{Jorgensen2015uniquenessconditions} to approximate the fraction of all gradient $s$-sparse vectors to uniquely solve \eqref{eq:ExactTV}.
The results prompt that there is a sharp transition between the case that every vector with a certain gradient sparsity is uniquely recoverable  and the case that TV-minimization will find a different solution than the desired vector. 
This behavior empirically agrees with the phase transition in the classical compressed sensing setup with Gaussian measurements \cite{donoho2004phasetransition}.

To efficiently check whether many specific vectors $x$ can be uniquely recovered via \eqref{eq:ExactTV}, one needs to establish characteristics of $x$ which must be easily verifiable.
Such a non-uniform recovery condition is given in the following theorem.

\begin{theo}\label{th:weakuniquenesscond}\cite{Jorgensen2015uniquenessconditions}
It holds that $x\in\R^n$ is a unique solution of \eqref{eq:ExactTV} if and only if there exists $w\in\R^m$ and $v\in\R^{n-1}$ such that
\begin{align}
\nabla^Tv = A^Tw, v_I = \mbox{sign}(\nabla x)_I, \Vert v_{I^c}\Vert_\infty < 1.\label{eq:uniquenesscondition}
\end{align}
\end{theo}
The basic idea of the proof is to use the optimality condition for convex optimization problems \cite{Rockafellar1972convex}.
Equivalent formulations of the latter theorem can be found in \cite{zhang2016uniquenessconditions,Kabanava2015uniquenessconditions} where the problem is considered from a geometric perspective.
However, verifying the conditions in Theorem \ref{th:weakuniquenesscond} still requires solving a linear program where an optimal $v$ for \eqref{eq:uniquenesscondition} needs to be found.
In classical compressed sensing, the \emph{Fuchs Condition} \cite{fuchs2004condition} is known as a weaker result as it suggests a particular $w$ in \eqref{eq:uniquenesscondition} and avoids solving the consequential linear program.
The following result generalizes this result to general analysis $\ell_1$-minimization.

\begin{coro}
If $x\in\R^n$ satisfies
\[\|(\identitymatrix_{I^c}^T\nabla(\nabla^T\identitymatrix_{I^c}\identitymatrix_{I^c}^T\nabla+A^TA)^{-1}\nabla\mbox{sign}(\nabla x))_I\|_\infty < 1\]
then $x$ is the unique solution of \eqref{eq:ExactTV}.
\end{coro}

\subsection{Recovery from Gaussian measurements}\label{sec:RecoveryGauss}
As discussed above, to date no deterministic constructions of compressed sensing matrices are known that get anywhere near an optimal number of measurements. Also for the variation of aiming to recover approximately gradient sparse measurements, the only near-optimal recovery guarantees have been established for random measurement models. Both under (approximate) sparsity and gradient sparsity assumptions, an important benchmark is that of a measurement matrix with independent standard Gaussian entries. Even though such measurements are hard to realize in practice, they can be interpreted as the scenario with maximal randomness, which often has particularly good recovery properties. For this reason, the recovery properties of total variation minimization have been analyzed in detail for such measurements. Interestingly, as shown by the following theorem, recovery properties in the one-dimensional case are significantly worse than for synthesis sparse signals and also for higher dimensional cases. That is why we focus on this case in Section~\ref{sec:RecoverySubgauss}, providing a geometric viewpoint and generalizing the results to subgaussian measurements.

\begin{theo}\cite{cai2015guarantees}\label{theo:Gauss1D}
	Let the entries of $A\in\R^{m\times n}$ be i.i.d. standard Gaussian random variables and let $\hat x$ be a solution of~\eqref{eq:ExactTV} with input data $y = Ax_0$. Then
	\begin{enumerate}
		\item There exist constants $c_1,c_2,c_3,c_4>0$, such that for $m\geq c_1\sqrt{sn}(\log n + c_2)$\[\Prob( \forall x_0\colon \|\nabla x_0\|_0\leq s \colon  \hat x = x_0)\geq 1-c_3\ee^{-c_4\sqrt{m}}.\]
		\item For any $\eta\in (0,1)$, there are constants $\tilde c_1, \tilde c_2>0$ and a universal constant $c_2>0$, such that for $s\geq \tilde c_0$ and $(s+1)<\frac n4$. If $m\leq \tilde c_1 \sqrt{sn}-\tilde c_2$, there exist infinitely many $x_0\in\R^n$ with $\|\nabla x_0\|_0\leq s$, such that $\Prob(\hat x \neq x_0)\geq 1-\eta$.
	\end{enumerate}
\end{theo}

This scaling is notably different from what is typically obtained for synthesis sparsity, where the number of measurements scales linearly with $s$ up to $\log$ factors. Such a scaling is only obtained for higher dimensional signals, e.g., images. Indeed, in~\cite{cai2015guarantees}, it is shown that for dimensions at least two the number of Gaussian measurements sufficient for recovery is \[m\geq \begin{cases}
c_2 s\log^3n,\,\,\text{ if }d=2\\
c_d s\log n,\,\,\text{ if }d\geq3,
\end{cases}
\]
where the constant $c_d$ depends on the dimension.

Furthermore, as we can see in Theorem~\ref{theo:Needell} below, this is also the scaling one obtains for dimensions larger than $1$ and Haar incoherent measurements. Thus the scaling of $\sqrt{sn}$ is a unique feature of the $1$-dimensional case.
Also note that the square-root factor in the upper bound makes the result meaningless for a sparsity level  on the order of the dimension. This has been addressed in \cite{KRZ15}, showing that a dimension reduction is also possible if the sparsity level is a (small) constant multiple of the dimension.

The proof of Theorem~\ref{theo:Gauss1D} uses Gordon's escape through the mesh Theorem~\cite{gordon1988milman}. We will elaborate on this topic in Section~\ref{sec:RecoverySubgauss}.

In case we are given noisy measurements $y = Ax_0 +e$ with $\| e \|_2\leq \varepsilon$, we can instead of solving~\eqref{eq:ExactTV} consider

\begin{equation}\label{eq:NoisyTV}\text{minimize } \| \nabla z\|_1\,\,\text{such that } \|Az -y\|_2\leq \varepsilon.\end{equation}

If $\nabla x_0$ is not exactly, but approximately sparse, and our measurements are corrupted with noise, the following result can be established.

\begin{theo}\label{theo:gaussNoisyComp}\cite{cai2015guarantees}
	Let the entries of $A\in\R^{m\times n}$ be i.i.d. standard Gaussian random variables and let $\hat x$ be a solution of~\eqref{eq:NoisyTV} with input data $y$ satisfying $\|Ax_0 - y\|_2\leq\varepsilon$. Then for any $\alpha\in(0,1)$, there are positive constants $\delta, c_0,c_1,c_2,c_3$, such that for $m = \alpha n$ and $s = \delta n$
	\[\Prob\left(\|x_0 - \hat x\|_2 \leq c_2 \frac{\min_{|S|\leq s} \| (\nabla x_0)_{S^c} \|_1}{\sqrt n} + c_3 \frac{\varepsilon}{\sqrt n}\right)\geq 1-c_0\ee^{-c_1 n}. \]
\end{theo}

This looks remarkably similar to the recovery guarantees obtained for compressed sensing, note however that the number of measurements needs to be proportional to $n$, which is not desirable. We will present a similar result with improved number of measurements in Section~\ref{subsec:Subgaussian}. 

\begin{theo}(Corollary of Theorem~\ref{theo:SubgaussGuarantee})
	Let $x_0\in\R^n$ be such that $\|\nabla x_0\|\leq s$ for $s>0$ and $A\in\R^{m\times n}$ with $m\geq C \sqrt{ns}\log(2n)$ be a standard Gaussian matrix. Furthermore, set $y = Ax_0+e$, where $\Vert e\Vert\leq \varepsilon$ denotes the (bounded) error of the measurement and for some absolute constants $c,\tilde c>0$
	the solution $\hat x$ of~\eqref{eq:minNoise} satisfies 
	\[\Prob\left(\|\hat x - x_0\| > \frac{2\eps}{c \sqrt[4]{ns}(\sqrt{\log(2n)}-1)}\right)\leq \ee^{-\tilde c \sqrt{ns}}.\]
\end{theo}

Note, however that in contrast to theorem~\ref{theo:gaussNoisyComp}, this theorem does not cover the case of gradient compressible vectors, but on the other hand Theorem~\ref{theo:SubgaussGuarantee} also incorporates the case of special subgaussian measurement ensembles. Also, if we set $s = \delta n$, we reach a similar conclusion as in Theorem~\ref{theo:gaussNoisyComp}. 

\subsection{Recovery from Haar-incoherent measurements}\label{sec:RecoveryHaarIncoherent}

For dimensions $d\geq 2$, Needell and Ward~\cite{needell2013near,needell2013stablereconsttv} derived recovery results for measurement matrices having the restricted isometry property (RIP) when composed with the Haar wavelet transform. Here we say that a matrix $\Phi$ has the RIP of order $k$ and level $\delta$ if for every $k$-sparse vector $x$ it holds that
\begin{equation*}
(1-\delta)\|x\|_2^2 \leq \|\Phi x\|_2^2 \leq (1-\delta)\|x\|_2^2.
\end{equation*}

 The results of~\cite{needell2013near,needell2013stablereconsttv} build upon a connection between a signal's wavelet representation and its total variation seminorm first noted by Cohen, Dahmen, Daubechies and DeVore~\cite{cohen2003harmonic}.

Their theorems yield stable recovery via TV minimization for $N^d$ dimensional signals. For $d=2$, notably these recovery results concern images of size $N\times N$.

Several definitions are necessary in order to be able to state the theorem. 
The $d$ dimensional discrete gradient is defined via $\nabla\colon \R^{C^d} \to \C^{N^d\times d}$ and maps $x\in\C^{N^d}$ to its discrete derivative which, for each $\alpha\in[N]^d$ is a vector $(\nabla x)_{\alpha}\in\C^d$ composed of the derivatives in all $d$ directions.
Up to now, we have always used the anisotropic version of the TV seminorm, which can be seen as taking the $\ell_1$ norm of the discrete gradient. The isotropic TV seminorm is defined via a combination of $\ell_2$ and $\ell_1$ norms. It is given by $\|z\|_{TV_2}:=\sum_{\alpha\in [N]^d} \|(\nabla z)_{\alpha}\|_2$. The result in~\cite{needell2013near} is given in terms of the isotropic TV seminorm but can also be formulated for the anisotropic version.

Furthermore, we will need to concatenate several measurement matrices in order to be able to state the theorem. This will be done via the concatenation operator $\oplus\colon \Lin(\C^n,\C^{k_1})\times\Lin(\C^n,\C^{k_2})\to \Lin(\C^n,\C^{k_1+k_2})$, which 'stacks' two linear maps.

Finally, we need the notion of shifted operators. For an operator $\Bcal\colon \C^{N^{l-1}\times (N-1)\times N^{d-l}}\to \C^q$, these are defined as the operators $\Bcal_{0_l}\colon\C^{Nˆd}\to\C^q$ and $\Bcal^{0_l}\colon\C^{Nˆd}\to\C^q$ concatenating  a column of zeros to the end or beginning of the $l$-th component, respectively.

\begin{theo}[\cite{needell2013near}]\label{theo:Needell}
	Let $N= 2^n$ and fix integers $p$ and $q$. Let $\Acal\colon \C^{N^d}\to\C^p$ be a map that has the restricted isometry property of order $2ds$ and level $\delta<1$ if it is composed with the orthonormal Haar wavelet transform. Furthermore let $\Bcal_1,\ldots,\Bcal_d$ with $\Bcal_j\colon \C^{(N-1)N^{d-1}}\to \C^q$ be such that $\Bcal = \Bcal_1\oplus\Bcal_2\oplus\cdots\oplus\Bcal_d$ has the restricted isometry property of order $5ds$ and level $\delta<\frac13$.
	Consider the linear operator $\Mcal = \Acal \oplus [\Bcal_1]_{0_1} \oplus [\Bcal_1]^{0_1}\oplus \cdots \oplus [\Bcal_d]_{0_d}\oplus [\Bcal_d]^{0_d}$. Then $\Mcal\colon \C^{N^d}\to \C^m$ with $m=2dq+p$ and for all $x\in \C^{N^d}$ we have the following.
	Suppose we have noisy measurements $y =\Mcal(x) +e$ with $\|e\|_2\leq\eps$, then the solution to 
	\[\hat x = \argmin_{z}\|z\|_{TV_2}\,\,\text{ such that } \|\Mcal(z) - y\|_2\leq\eps\]
	satisfies
	\begin{enumerate}
		\item $\|\nabla(x-\hat x) \|_2\leq c_1 \left( \frac{\|\nabla x -(\nabla x)_S\|_{1,2}}{\sqrt s} +\sqrt d \eps\right)$,
		\item $\|x-\hat x\|_{TV_2}\leq c_2\left(\|\nabla x - (\nabla x)_S\|_{1,2} +\sqrt{sd} \eps \right)$,
		\item $\|x - \hat x\|_2 \leq c_3 d\log N \left( \frac{\|\nabla x -(\nabla x)_S\|_{1,2}}{\sqrt s} +\sqrt d \eps \right),$
	\end{enumerate}
	for some absolute constants $c_1,c_2,c_3$.
\end{theo}

From the last point of the previous theorem, we see that for noiseless measurements and gradient sparse vectors $x$, perfect recovery can be achieved provided the RIP assumption holds. Subgaussian measurement matrices, for example, will have the RIP, also when composed with the Haar wavelet transform $H$ (this is a direct consequence of rotation invariance). Moreover, as shown in \cite{KW11}, randomizing the column signs of an RIP matrix will, with high probability, also yield a matrix that has the RIP when composed with $H$. An important example is a subsampled Fourier matrix with random column signs, which relates to spread spectrum MRI (cf.~\cite{PMGTVVW12}).
\subsection{Recovery from subsampled Fourier measurements}
Fourier measurements are widely used in many applications.
Especially in medical applications as parallel-beam tomography and magnetic resonance imaging it is desirable to reduce the number of samples to spare patients burden.
In Section \ref{suffcond}, this is a motivation for introducing algorithmic checks for unique solutions of \eqref{eq:ExactTV}. In this section, we consider a probabilistic approach where an incomplete measurement matrix $A\in\mathbb{C}^{m\times n}$ chosen from the discrete Fourier transform on $\mathbb{C}^N$ is considered.
Therefore we consider a subset $\Omega$ of the index set $\{-\lfloor n/2\rfloor+1,...,\lceil n/2\rceil\}$, where $\Omega$ consists of $m$ integers chosen uniformly at random and, additionally, $0\in\Omega$. 
Hence, we want to recover a signal, sparse in the gradient domain, with a measurement matrix $A = (e^{2\pi ikj/n})_{k\in\Omega,j\in[n]}$.
In \cite{crt2006robust} the optimal sampling cardinality for $s$-sparse signals in the gradient domain was given and enables to recover one-dimensional signals signals from $\mathcal{O}(k\log(n))$ Fourier samples.
It naturally extends to two dimensions.

\begin{theo}\cite{crt2006robust}
With probability exceeding $1-\eta$, a signal $z$, which is $k$-sparse in the gradient domain is the unique solution of \eqref{eq:ExactTV} if \[m\gtrsim k(\log(n)+\log(\eta^{-1})).\]
\end{theo}
As already discussed in the introduction, the proof of this result proceeds via  recovering the gradient and then using that the discrete gradient (with periodic boundary conditions) is injective. Due to the poor conditioning of the gradient, however, this injectivity results do not directly generalize to recovery guarantees for noisy measurements. For two (and more) dimensions, such results can be obtained via the techniques discussed in the previous subsection. 

These techniques, however, do not apply directly. Namely, the Fourier (measurement) basis is not incoherent to the Haar wavelet basis; in fact, the constant vector is contained in both, which makes them maximally coherent. As observed in \cite{pvw11}, this incoherence phenomenon only occurs for low frequencies, the high frequency Fourier basis vectors exhibit small inner products to the Haar wavelet basis. This can be taken into account using a {\em variable density} sampling scheme with sampling density that is larger for low frequencies and smaller for high frequencies. For such a sampling density, one can establish the restricted isometry for the corresponding randomly subsampled discrete Fourier matrix combined with the Haar wavelet transform with appropriately rescaled rows \cite{KW14}. This yields the following recovery guarantee.

\begin{theo}\cite{KW14}
	\label{thm1}
	Fix integers $N = 2^p, m,$ and $s$ such that $s\gtrsim \log(N)$ and 
	\begin{equation}
	\label{m:tv}
	m \gtrsim s \log^{3}(s)\log^5(N).
	\end{equation}
	Select $m$ frequencies $\{ (\omega_1^j, \omega_2^j ) \}_{j=1}^m \subset \{-N/2+1, \dots, N/2\}^2$ i.i.d. according to
	\begin{equation}
	\label{inverseD}
	\mathbb{P} \big[ (\omega_1^j, \omega_2^j) = (k_1, k_2) \big]  = C_N \min\left(C, \frac{1}{k_1^2 + k_2^2 }\right) =: \eta(k_1, k_2), \quad -N/2+1 \leq k_1, k_2 \leq N/2,
	\end{equation}
	where $C$ is an absolute constant and $C_N$ is chosen such that $\eta$ is a probability distribution. \\
	Consider the weight vector $\rho = (\rho_j)_{j=1}^m$ with $\rho_j = (1/\eta(\omega^j_1, \omega^j_2))^{1/2}$, and assume that the noise vector $\xi = (\xi_j)_{j=1}^m$ satisfies $\|\rho \hspace{.5mm} \circ \hspace{.5mm}  \xi\|_2\leq  \varepsilon \sqrt{m} $, for some $\epsilon>0$.  Then with probability exceeding $1 - N^{-C\log^3(s)}$, the following holds for all images $f \in \C^{N \times N}$:
	
	\noindent Given noisy partial Fourier measurements $y = {\cal F}_{\Omega}f + \xi$, the estimation
	\begin{equation}
	\label{TV}
	f^{\#} = \argmin_{g \in \C^{N \times N}} \| g \|_{TV}  \quad  \textrm{such that}  \quad \|\rho\circ( {\cal F}_{\Omega} g - y) \|_2   \leq \varepsilon\sqrt{m},
	\end{equation}
	where $\circ$ denotes the Hadamard product, approximates $f$ up to the noise level and best $s$-term approximation error of  its gradient:
	\begin{equation}
	\label{stable3}
	\| f -f^{\#} \|_2 \lesssim  \frac{ \| \nabla f -  (\nabla f)_s \|_1}{\sqrt{s}}+ \varepsilon.
	\end{equation}
\end{theo}

A similar optimality result is given in \cite{poon2013roleoftv}, also for noisy data  and inexact sparsity. In contrast to the previous result, this result includes the one-dimensional case.
The key to obtaining such a result is showing that the stable gradient recover implies the stable signal recovery, i.e.,
\begin{align}
\|z\|_2\lesssim\gamma+\Vert z\Vert_{TV}\mbox{ with }\|Az\|_2\le\gamma.\label{eq:sobolevemb}
\end{align}
Again the sampling distribution is chosen as a combination of the uniform distribution and a decaying distribution.
The main idea is to use this sampling to establish \eqref{eq:sobolevemb} via the RIP.
We skip technicalities for achieving the optimality in the following theorem and refer to the original article for more details. 

\begin{theo}\cite{poon2013roleoftv}
Let $z\in\mathbb{C}^n$ be fixed and $x$ be a minimizer of \eqref{eq:NoisyTV} with $\eps = \sqrt{m}\delta$ for some $\delta>0$, $m\gtrsim k\log(n)(1+\log(\eta^{-1}))$, and an appropriate sampling distribution.
Then with probability exceeding $1-\eta$, it holds that
\[\|\nabla z-\nabla x\|_2\lesssim\left(\delta\sqrt{k}+C_1\frac{\|P\nabla z\|_1}{\sqrt{k}}\right), \frac{\|z-x\|_2}{\sqrt{n}}\lesssim C_2\left(\frac{\delta}{\sqrt{s}}+C_1\frac{\|P\nabla z\|_1}k\right),\]
where $P$ is the orthogonal projection onto a $k$-dimensional subspace, 
\[C_1 = \log(k)\log^{1/2}(m)\mbox{, and }C_2 = \log^2(k)\log(n)\log(m).\] 
\end{theo}

In the two-dimensional setting the result changes to 
\[\|\nabla z-\nabla x\|_2\lesssim\left(\delta\sqrt{k}+C_3\frac{\|P\nabla z\|_1}{\sqrt{k}}\right), \|z-x\|_2\lesssim C_2\left(\delta+C_3\frac{\|P\nabla z\|_1}k\right),\]
with remaining $C_2$ and
\[C_3 = \log(k)\log(n^2/k)\log^{1/2}(n)\log^{1/2}(m).\]
These results are optimal since the best error one can archive \cite{needell2013stablereconsttv} is $\|z-x\|_2\lesssim \|P\nabla z\|_1k^{-1/2}$.

The optimality in the latter theorems is achieved by considering a combination of uniform random samling and variable density sampling.
Uniform sampling on its own can achieve robust and stable recovery.
However, the following theorem shows that the signal error is no longer optimal but the bound on the gradient error is still optimal up to log factors.
Here \eqref{eq:sobolevemb} is obtained by using the Poincar\'e inequality.

\begin{theo}\cite{poon2013roleoftv}
Let $z\in\mathbb{C}^n$ be fix and $x$ be a minimizer of \eqref{eq:NoisyTV} with $\eps = \sqrt{m}\delta$ for some $\delta>0$ and $m\gtrsim k\log(n)(1+\log(\eta^{-1}))$ with random uniform sampling.
Then with probability exceeding $1-\eta$, it holds that
\[\|\nabla z-\nabla x\|_2\lesssim\left(\delta\sqrt{k}+C\frac{\|P\nabla z\|_1}{\sqrt{k}}\right), \frac{\|z-x\|_2}{\sqrt{n}}\lesssim(\delta\sqrt{s}+C\|P\nabla z\|_1),\]
where $P$ is the orthogonal projection onto a $k$-dimensional subspace and $C = \log(k)\log^{1/2}(m)$.
\end{theo}
%


%

\section{TV-recovery from subgaussian measurements in 1D}\label{sec:RecoverySubgauss}

In this section, we will apply the geometric viewpoint discussed in~\cite{vershynin2014estimation} to the problem, which will eventually allow us to show the TV recovery results for noisy subgaussian measurements mentioned in Section~\ref{sec:RecoveryGauss}.

As in the original proof of the 1D recovery guarantees for Gaussian measurements \cite{cai2015guarantees}, the {\em Gaussian mean width} will play an important role in our considerations.
{\defi The (Gaussian) mean width of a bounded subset $K$ of $\R^n$ is defined as \[w(K) := \Ex \sup_{x\in K-K}\ip{g}{x},\]
	where $g\in\R^n$ is a vector of i.i.d. $\Ncal(0,1)$ random variables.
}
 
In \cite{cai2015guarantees}, the mean width appears in the context of the {\em Gordon's escape through the mesh} approach \cite{gordon1988milman} (see Section~\ref{sec:ExactRecov} below), but as we will see, it will also be a crucial ingredient in applying the Mendelson small ball method \cite{koltchinskii2015bounding, mendelson2014learning}.

The mean width has some nice (and important) properties, it is for example invariant under taking the convex hull, i.e.,
\[w(\ch(K))  = w(K).\]
Furthermore, it is also invariant under translations of $K$, as $(K-x_0) - (K-x_0) = K-K$.  Due to the rotational invariance of Gaussian random variables, that is $Ug\sim g$, we also have that $w(UK) = w(K)$.
Also, it satisfies the inequalities 
\[w(K) = \Ex \sup_{x\in K-K}\ip{g}{x}\leq 2\Ex\sup_{x\in K}\ip{g}{x}\leq 2\Ex\sup_{x\in K}|\ip{g}{x}|,\]
which are equalities if $K$ is symmetric about $0$, because then $K=-K$ and hence $K-K=2K$.

\subsection{$M^*$ bounds and recovery}
In order to highlight the importance of the Gaussian mean width in signal recovery, we present some arguments from \cite{vershynin2014estimation}.
Thus in this section we present a classical result, the $M^*$ bound, which connects the mean width to recovery problems, cf.~\cite{vershynin2014estimation}. 
Namely, recall that due to rotational invariance, the kernel of a Gaussian random matrix $A\in\R^{m\times n}$ is a random subspace distributed according to the uniform distribution (the Haar measure) on the Grassmannian 

\[G_{n,n-m}:=\{V\leq \R^n\colon \dim(V) = n-m\}.\]

Consequently, the set of all vectors that yield the same measurements directly correspond to such a random subspace.

 The average size of the intersection of this subspace with a set reflecting the minimization objective now gives us an average bound on the worst case error.

{\theo[$M^*$ bound, Theorem~3.12 in \cite{vershynin2014estimation}]\label{theo:MStarBound} Let $K$ be a bounded subset of $\R^n$ and $E$ be a random subspace of $\R^n$ of drawn from the Grassmanian $G_{n,n-m}$ according to the Haar measure. Then
	\begin{equation}\label{M*bound}
	\Ex\, \diam(K\cap E) \leq C \frac{w(K)}{\sqrt{m}},
	\end{equation}
where $C$ is absolute constant.	
}

Given the $M^*$-bound it is now straightforward to derive bounds on reconstructions from linear observations. We first look at feasibility programs - which in turn can be used to obtain recovery results for optimization problems.
For that, let $K\subset \R^n$ be bounded and $x\in K$ be the vector we seek to reconstruct from measurements $Ax = y$ with a Gaussian matrix $A\in\R^{m\times n}$.

{\coro\label{theo:feasibility}\cite{mendelson2007reconstruction} Choose $\hat x\in \R^n$, such that 
	\[\hat x \in K\,\text{ and } A\hat x = y,\]
	then one has, for an absolute constant $C'$,
	\[\Ex \sup_{x\in K}\Vert \hat x - x\Vert_2 \leq C' \frac{w(K)}{\sqrt{m}}.\]
}

This corollary directly follows by choosing $C'=2C$, observing that $\hat x - x\in K-K$, and that the side constraint enforces $A(\hat x - x) = 0$.

Via a standard construction in functional analysis, the so called \emph{Minkowski functional}, one can now cast an optimization problem as a feasiblity program so that Corollary~\ref{theo:feasibility} applies.

{\defi The Minkowski functional of a bounded, symmetric set $K\subset \R^n$ is given by 
	\[\Vert\cdot\Vert_K\colon \R^n \to \R\colon x\mapsto \inf\{t>0\colon x\in tK\}.\]
}

So the Minkowski functional tells us, how much we have to 'inflate' our given set $K$ in order to capture the vector $x$. Clearly, from the definition we have that if $K$ is closed
\[K = \{x\colon \Vert x\Vert_K\leq 1\}.\]
If a convex set $K$ is closed and symmetric, then $\Vert\cdot\Vert_K$ defines a norm on $\R^n$.

Recall that a set $K$ is star shaped, if there exists a point $x_0\in K$, which satisfies that for all $x\in K$ we have $\{tx_0+(1-t)x\colon t\in[0,1]\}\subset K$.
It is easy to see that convex sets are star shaped, but for example unions of subspaces are not convex, but star shaped.

For bounded, star shaped $K$, the notion of $\Vert \cdot\Vert_K$ now allows to establish a direct correspondence between norm minimization problems and feasibility problems. With this observation, Corollary~\ref{theo:feasibility} translates to the following result.

{\coro\label{theo:opti} For $K$ bounded, symmetric and star-shaped, let $x\in K$ and $y=Ax$. Choose $\hat x\in \R^n$, such that it solves
	\[\min\Vert z\Vert_K \,\text{ with } Az = y,\]
	then
	\[\Ex \sup_{x\in K}\Vert \hat x - x\Vert_2 \leq C' \frac{w(K)}{\sqrt{m}}.\]
}

Here $\hat x\in K$ is due to the fact that the minimum satisfies $\Vert \hat x\Vert_K \leq \Vert x \Vert_K\leq 1$, as $x\in K$ by assumption. 

This result directly relates recovery guarantees to the mean width, it thus remains to calculate the mean width for the sets under consideration. In the following subsections, we will discuss two cases. The first one directly corresponds to the desired signal model, namely gradient sparse vectors. These considerations are mainly of theoretical interest, as the associated minimization problem closely relates to support size minimization, which is known to be NP hard in general. The second case considers the TV minimization problem introduced above, which then also yields guarantees for the (larger) set of vectors with bounded total variation.

Note, however, that the $M^*$-bound only gives a bound for the expected error. We can relate this result to a statement about tail probabilities using Markov's inequality, namely \[\Prob(\sup_{x\in K} \|x-\hat x\|_2>t)\leq t^{-1}\Ex \sup_{x\in K} \|x-\hat x\|_2\leq C'\frac{w(K)}{t\sqrt{m}}.\]

In the next section we compute the mean width for the set of gradient sparse vectors, that is we now specify the set $K$ in Corollary~\ref{theo:feasibility} to be the set of all vectors with energy bounded by one that only have a small number of jumps.

\subsection{The mean width of gradient sparse vectors in 1d}\label{sec:GradSparse}

Here~\cite{plan2013robust} served as an inspiration, as the computation is very similar for the set of sparse vectors.

{\defi The jump support of a vector $x$ is given via
	\[\tvsupp(x):= \{i\in[n-1]\colon x_{i+1}-x_i\neq 0\}.\]
}
The jump support captures the positions, in which a vector $x$ changes its values. With this, we now define the set 
\[K_0^s:= \{x\in\R^n\colon\Vert x\Vert_2\leq 1, |\tvsupp(x)|\leq s\}.\]
The set $K_0^s$ consists of all $s$-gradient sparse vectors, which have $2$-norm smaller than one. We will now calculate the mean width of $K_0^s$ in order to apply Corrolary~\ref{theo:feasibility} or~\ref{theo:opti}.

Note that we can decompose the set $K_0^s$ into smaller sets $K_J\cap B_2^n$ with $K_J=\{x\colon \tvsupp(x)\subset J\}$, $|J| = s$ and $B_2^n=\{x\in \R^n\colon \Vert x\Vert_2\leq 1\}$. As we can't add any jumps within the set $K_J$, it is a subspace of $\R^n$. We can even quite easily find an orthonormal basis for it, if we define
\[(e_{[i,j]})_k:= \frac1{\sqrt{j-i+1}}\begin{cases}1,\, \text{ if } k\in[i,j]\\ 0,\,\,\text{else}\end{cases}.\]
As we can align all elements of $J = \{j_1,j_2,\ldots, j_s\}$ with $1\leq j_1<j_2<\ldots < j_s =n$, we see that $\{e_{[1,j_1]},e_{[j_1+1,j_2]},e_{[j_2+1,j_3]},\ldots,e_{[j_{s-1}+1,j_s]}\}$ forms an ONB of $K_J$.
Now, we can write all elements $x\in K_J\cap B_2^n$ as $x = \sum_{i=1}^s \alpha_i e_{[j_{i-1}+1,j_i]}$ by setting $j_0:=0$. The property that $x\in B_2^n$ now enforces (ONB) that $\Vert \alpha \Vert_2\leq 1$.  Now, note that $K_0^s = -K_0^s$, so we have 
\[w(K_0^s) = \Ex \sup_{x\in K_0^s-K_0^s}\ip{g}{x} = 2\Ex \sup_{x\in K_0^s}\ip{g}{x}.\]

Using the decomposition $K_0^s = \bigcup_{|J| = s}\left(K_J\cap B_2^n\right)$, we get
\[w(K_0^s)  =  2\Ex \sup_{|J|= s}\sup_{x\in K_J\cap B_2^n}\ip{g}{x}.\]
Now 
\[\sup_{x\in K_J\cap B_2^n}\ip{g}{x} \leq \sup_{\alpha\in B_2^s}\sum_{i=1}^s \alpha_i\ip{g}{e_{[j_{i-1}+1,j_i]}}= \sup_{\alpha\in B_2^s}\sum_{i=1}^s \alpha_i\underbrace{\sum_{k = j_{i-1}+1}^{j_i} \frac{g_k}{\sqrt{j_i-j_{i-1}}}}_{=:G_i^J}.\]
Note that $G_i^J$ is again a Gaussian random variable with mean $0$ and variance $1$. Furthermore, the supremum over $\alpha$ is attained, if $\alpha$ is parallel to $G^J$, so we have
$\sup_{x\in K_J\cap B_2^n}\ip{g}{x} = \Vert G^J\Vert_2$. Also note that $G^J$ has i.i.d. entries, but for different $J_1,J_2$, the random vectors $G^{J_1}$ and $G^{J_2}$ may be dependent.
Our task is now to calculate $\Ex\sup_{|J|=s}\Vert G^J\Vert_2$.
As it has been shown for example in~\cite{foucart2013mathematical}, we have that
\[\sqrt{\frac2\pi}\sqrt{s}\leq \Ex\Vert G^J\Vert_2\leq \sqrt{s}\]
and from standard results for Gaussian concentration (cf.~\cite{plan2013robust}), we get
\[\Prob(\Vert G^J\Vert_2\geq\sqrt{s} + t)\leq \Prob(\Vert G^J\Vert_2\geq\Ex\Vert G_J\Vert_2 + t)\leq \ee^{-t^2/2}.\]
By noting that $|\{J\subset[n]\colon |J|=s\}| = { n\choose s}$, we see by a union bound that

\[
\Prob(\sup_{|J|=s}\Vert G^J\Vert_2\geq\sqrt{s} + t) \leq {n\choose s}\Prob(\Vert G^J\Vert_2\geq\sqrt{s} + t)\leq {n\choose s}\ee^{-t^2/2}.
\]

For the following calculation, set $X:= \sup_{|J|=s}\Vert G^J\Vert_2$. By Jensen's inequality and rewriting the expectation, we have that
\[\ee^{\lambda \Ex X}\leq \Ex \ee^{\lambda X} = \int_0^\infty \Prob(\ee^{\lambda X}\geq \tau)\dx \tau.\]
Now, the previous consideration showed, that 
\[\Prob(\ee^{\lambda X}\geq\underbrace{\ee^{\lambda (\sqrt{s} + t)}}_{=:\tau}) = \Prob(X\geq\sqrt{s} + t)\leq  {n\choose s}\ee^{-t^2/2}=  {n\choose s}\ee^{-(\log(\tau)/\lambda -\sqrt{s})^2/2},
\]
Computing the resulting integrals yields 
\begin{equation*}
\ee^{\lambda \Ex X}\leq{n\choose s} \ee^{-s/2}\lambda\sqrt{2\pi}\ee^{(\sqrt{s}+\lambda)^2/2}.
\end{equation*}

Using a standard bound for the binomial coefficients, namely ${n\choose s}\leq \ee^{s \log (\ee n/s)}$, we see

\[\ee^{\lambda \Ex X}\leq \ee^{s\log(\ee n/s) - s/2 +(\sqrt{s} + \lambda)^2/2+\log(\lambda)+\log(\sqrt{2\pi})},\]
or equivalently
\[\lambda \Ex X \leq s\log(\ee n/s) - s/2 +(\sqrt{s} + \lambda)^2/2+\log(\lambda)+\log(\sqrt{2\pi})\]
By setting $\lambda = \sqrt{s\log(\ee n/s) }$ and assuming (reasonably) large $n$, we thus get
\[\Ex X\leq 5 \sqrt{s\log(\ee n/s) }.\]
From this, we see that
\[w(K_0^s)\leq 10\sqrt{s\log(\ee n/s) }.\]

It follows that the Gaussian mean width of the set of gradient sparse vectors is the same as the mean width of sparse vectors due to the similar structure.
If we want to obtain accuracy $\delta$ for our reconstruction, according to Theorem~\ref{theo:feasibility}, we need to take
\[m = \Ocal\left(\frac{s\log(\ee n/s)}{\delta^2}\right)\] 
measurements.

In Compressed Sensing, the squared mean width of the set of $s$-sparse vectors (its so called \emph{statistical dimension}) already determines the number of required measurements in order to recover a sparse signal with basis pursuit. This is the case because the convex hull of the set of sparse vectors can be embedded into the $\ell_1$-ball inflated by a constant factor.
In the case of TV minimization, as we will see in the following section, this embedding yields a (rather large) constant depending on the dimension.
 
\subsection{The extension to gradient compressible vectors needs a new approach}

In the previous subsection, we considered exactly gradient sparse vectors. However searching all such vectors $x$ that satisfy $Ax=y$ is certainly not a feasible task. Instead, we want to solve the convex program

\[\min\Vert z\Vert_{TV} \,\text{ with } Az = y,\]
with $\Vert z\Vert_{TV} = \Vert \nabla z\Vert_1$ the total variation seminorm. Now if we have that $x\in K_0^s$, we get that
\[\Vert x\Vert_{TV}  \leq 2\Vert \alpha\Vert_1 \leq 2\sqrt{s}\Vert\alpha\Vert_2 = 2\sqrt{s},\]
with $\alpha$ as in section~\ref{sec:GradSparse}, so $K_0^s\subset K_{TV}^{2\sqrt{s}}:=\{x\in B_2^n\colon \Vert x\Vert_{TV}\leq 2\sqrt{s}\}$. As $K_{TV}^{2\sqrt{s}}$ is convex, we even have $\ch(K_0^s)\subset K_{TV}^{2\sqrt{s}}$.
We can think of the set $K_{TV}^{2\sqrt{s}}$ as 'gradient- compressible' vectors.

In the proof of Theorem 3.3 in~\cite{cai2015guarantees}, the Gaussian width of the set $K_{TV}^{4\sqrt{s}}$ has been calculated via a wavelet-based argument. One obtains that $w(\Ktv)\leq C\sqrt{\sqrt{ns}\log(2n})$ with $C\leq20$ being an absolute constant. In this section we illustrate, that proof techniques different from the ones used in the case of synthesis sparsity are indeed necessary in order to obtain useful results. In the synthesis case, the $1$-norm ball of radius $\sqrt s$ is contained in the set of $s$-sparse vectors inflated by a constant factor. This in turn implies that the mean width of the compressible vectors is bounded by a constant times the mean width of the $s$-sparse vectors.

We will attempt a similar computation, that is to find a constant, such that the set $K_{TV}^{2\sqrt{s}}$ is contained in the 'inflated' set $c_{n,s} \ch(K_0^s)$. Then $w(\Ktv)\leq c_{n,s} w(K_0^s)$. Although this technique works well for sparse recovery, where $c_{n,s} = 2$, it pityably fails in the case of TV recovery as we will see below.

Let us start with $x\in \Ktv$. Now we can decompose $J:=\tvsupp(x) = J_1\uplus J_2\uplus\ldots J_p$ with $|J_k|\leq s$ in an ascending manner, i.e., for all $k\in J_i, l\in J_{i+1}$, we have that $\alpha_k<\alpha_l$. Note that the number $p$ of such sets satisfies $p\leq \frac{n}{s}$.
Similarly as above, we now write $x = \sum_{i=1}^{|J|}\alpha_i e_{[j_{i-1}+1,j_i]} = \sum_{k=1}^p\sum_{i\in J_k}\alpha_i e_{[j_{i-1}+1,j_i]}$. From this, we see that 
\[x =  \sum_{k = 1}^p \Vert \alpha_{J_k}\Vert_2 \underbrace{\sum_{i\in J_k}\frac{\alpha_i}{\Vert \alpha_{J_k}\Vert_2} e_{[j_{i-1}+1,j_i]}}_{\in K_0^s}.\]
The necessary factor $c_{n,s}$ can be found by bounding the size of $\Vert \alpha_{J_k}\Vert_2$, namely
\[\max (\Vert \alpha_{J_k}\Vert_2)\leq \sum_{k=1}^p\Vert \alpha_{J_k}\Vert_2\stackrel{C-S}{\leq} \underbrace{\Vert \alpha\Vert_2}_{\leq 1} \sqrt{p}\leq\sqrt{\frac{n}{s}}.\]

From this, we see that $\Ktv\subset \sqrt{\frac{n}{s}} \ch(K_0^s)$. To see that this embedding constant is optimal, we construct a vector, for which it is needed.

To simplify the discussion, suppose that $n$ and $s$ are even and $s|n$. 
For even $n$, the vector $x_1 = (\sqrt{\frac{1-(-1)^k\varepsilon}{n}})_k$ has unity norm, lies in $\Ktv$ for $\varepsilon<\frac{2\sqrt{s}}{n}$ and has jump support on all of $[n]$!

For a vector $x\in \R^n$ and an index set $I\subset [n]$, we define the restriction of $x$ to $I$ by \[(x|_I)_j :=\begin{cases}x_j\text{, if }j\in I\\
0\text{, else}.\end{cases}\]
By splitting $\tvsupp(x_1)$ into sets $J_1,\ldots,J_{n/s}$ and setting $a_k = \sqrt{\frac ns}x_1|_{J_k}\in K_0^s$, we see that $x_1= \sum_{k=1}^{n/s} \sqrt{\frac sn} a_k$ and in order for this to be elements of $c_{n,s}\ch(K_0^s)$, we have to set $c_{n,s} = \sqrt{\frac ns}$. This follows from
\[x_1= \sum_{k=1}^{n/s} x_1|_{J_k}=\sum_{k=1}^{n/s} \sqrt{\frac sn} \frac{p}{p}a_k =  \sum_{k=1}^{n/s} \frac{1}{p} \underbrace{\left(\sqrt{\frac ns}a_k\right)}_{\in \sqrt{\frac ns} K_0^s}\in \sqrt{\frac ns}\ch(K_0^s)\]
and no smaller inflation factor than $\sqrt{\frac ns}$ can suffice.\\

So from the previous discussion, we get
{\lemma\label{lemma:inclusions} We have the series of inclusions
	\[\ch(K_0^s)\subset \Ktv \subset \sqrt{\frac ns}\ch(K_0^s).\]}

In view of the results obtainable for sparse vectors and the $\ell_1$-ball, this is very disappointing, because Lemma~\ref{lemma:inclusions} now implies that the width of $\Ktv$ satisfies 
\[w(\Ktv)\leq w\left(\sqrt{\frac ns}\ch(K_0^s)\right) = \sqrt{\frac ns} w(K_0^s)\leq 10\sqrt{n\log(\ee (n-1)/s)},\]
which is highly suboptimal. 

Luckily, the results in~\cite{cai2015guarantees} suggest, that the factor $n$ in the previous equation can be replaced by $\sqrt{sn}$. However, they have to resort to a direct calculation of the Gaussian width of $\Ktv$. The intuition why the Gaussian mean width can be significantly smaller than the bound given in Lemma~\ref{lemma:inclusions} stems from the fact, that in order to obtain an inclusion we need to capure all 'outliers' of the set - no matter how small their measure is.

\subsection{Exact recovery}\label{sec:ExactRecov}

 For exact recovery, the $M^*$-bound is not suitable anymore and, as suggested in~\cite{vershynin2014estimation}, we will use 'Gordon's escape through the mesh' in order to find conditions on exact recovery. Exact recovery for TV minimization via this approach has first been considered in~\cite{cai2015guarantees}. 


Suppose, we want to recover $x\in K_0^s$ from Gaussian measurements $Ax =y$. Given, that we want our estimator $\hat x$ to lie in a set $K$, exact recovery is achieved, if $K\cap \{z\colon Az = y\} = \{x\}$. This is equivalent to requiring
\[(K-x)\cap  \underbrace{\{z-x \colon Az = y\}}_{=\ker(A)}  = \{0\}.\]
With the descent cone $D(K,x) = \{t(z-x)\colon t\geq0,z\in K\}$, we can rewrite this condition as 
\[D(K,x)\cap  \ker(A) = \{0\},\]
by introducing the set $S(K,x) = D(K,x)\cap B_2^n$, we see that if
\[S(K,x)\cap  \ker(A) =\emptyset,\]
we get exact recovery.
The question, when a section of a subset of the sphere with a random hyperplane is empty is answered by Gordon's escape through a mesh.
{\theo[\cite{gordon1988milman}] Let $S\subset \S^{n-1}$ be fixed and $E\in G_{n,n-m}$ be drawn at random according to the Haar measure. Assume that $\hat w(S)=\Ex \sup_{u\in S}\ip{g}{u}<\sqrt{m}$, then $S\cap E=\emptyset$ with probability exceeding 
	\[1-2.5\exp\left(-\frac{(m/\sqrt{m+1}-\hat w(S))^2}{18}\right).\]
}

So we get exact recovery with high probability from a program given in Theorem~\ref{theo:feasibility} or~\ref{theo:opti}, provided that $m>\hat w(S(K,x_0))^2$.

Let's see how this applies to TV minimization. Suppose, we are given $x\in K_0^s$ and Gaussian measurements $Ax=y$. Solving
\[\min\Vert z\Vert_{TV} \,\text{ with } Az = y,\]
amounts to using the Minkowski functional of the set $K = \{z\in\R^n\colon \Vert z\Vert_{TV}\leq \Vert x\Vert_{TV}\}$, which is a scaled TV-Ball.

In~\cite{cai2015guarantees}, the null space property for TV minimization given in Corollary~\ref{corr:TVNSP} has been used in order to obtain recovery guarantees.

They consider the set, where this condition is not met
\[\Scal := \{x'\in B_2^n\colon \exists J \subset[n],|J|\leq s, \Vert(\nabla x')_J\Vert_1 \geq \Vert (\nabla x')_{J^c}\Vert_1\},\]
and apply Gordon's escape through the mesh to see that with high probability, its intersection with the kernel of $A$ is empty, thus proving exact recovery with high probability. Their estimate to the mean width of the set $\Scal$, 
\[\hat w(\Scal)\leq c \sqrt[4]{ns}\sqrt{\log(2n)}\]
with $c<19$ is essentially optimal (up to logarithmic factors), as they also show that $w(\Scal)\geq C \sqrt[4]{ns}$.
So uniform exact recovery can only be expected for $m =\Ocal(\sqrt{sn}\log n)$ measurements.

Let us examine some connections to the previous discussion about the descent cone.

{\lemma We have that for $K = \{z\in\R^n\colon \Vert z\Vert_{TV}\leq \Vert x\Vert_{TV}\}$ defined as above and $x\in K_0^s$, it holds that $S(K,x)\subset \Scal$.
}
\begin{proof}
	Let $y\in S(K,x)$. Then there exists a $x\neq z\in K$, such that $y = \frac{z-x}{\Vert z-x\Vert_2}$. Set $J=\tvsupp(x)$, then, as $z\in K$, we have that $\Vert z\Vert_{TV}\leq \Vert x\Vert_{TV}$, or
	\[\sum_{i\in J} |(\nabla x)_i|\geq \sum_{i\in J} |(\nabla z)_i|+\sum_{i\not\in J} |(\nabla z)_i|\]
	Now, by the triangle inequality and this observation, we have
	\[\sum_{i\in J} |(\nabla x)_i- (\nabla z)_i|\geq \sum_{i\in J} |(\nabla x)_i|- |(\nabla z)_i|\geq \sum_{i\not\in J} |(\nabla z)_i| = \sum_{i\not\in J} |(\nabla x)_i- (\nabla z)_i|.\]
	The last equality follows from the fact that $\nabla x$ is zero outside of the gradient support of $x$. Multiplying both sides with $\frac1{\Vert z-x\Vert_2}$ gives the desired result
	\[\Vert(\nabla y)_J\Vert_1 =\frac1{\Vert z-x\Vert_2} \sum_{i\in J} |(\nabla x)_i- (\nabla z)_i|\geq\\ \geq\frac1{\Vert z-x\Vert_2} \sum_{i\not\in J} |(\nabla x)_i- (\nabla z)_i|=\Vert(\nabla y)_{J^c}\Vert_1. \]
	
\end{proof}

The previous lemma shows that the recovery guarantees derived from the null space property and via the descent cone are actually connected in a very simple way.

Clearly, now if we do not intersect the set $\Scal$, we also do not intersect the set $S(K,x)$, which yields exact recovery for example with the same upper bounds on $m$ as for $\Scal$. 
Even more specifically, in the calculation of $\hat w(\Scal)$ given in~\cite{cai2015guarantees}, an embedding into a slightly larger set $\tilde \Scal = \{x\in B_2^n\colon \Vert x\Vert_{TV}\leq 4\sqrt{s}\}$ is made. This embedding can also quite easily be done if we note that $\Vert x\Vert_{TV}\leq 2\sqrt{s}$, as we showed above and $\Vert z\Vert_{TV}\leq \Vert x\Vert_{TV}$.

Note that the same discussion also holds for higher dimensional signals, such that the improved numbers of measurements as given in Section~\ref{sec:RecoveryGauss} can be applied.

\subsection{Subgaussian measurements}\label{subsec:Subgaussian}

Up to this point, all our measurement matrices have been assumed to consist of i.i.d.~Gaussian random variables. We will reduce this requirement in this section to be able to incorporate also subgaussian measurement matrices into our framework. 

\begin{defi}\label{defi:subgauss}
	A real valued random variable $X$ is called \emph{subgaussian}, if there exists a number $t>0$, such that $\Ex \ee^{tX^2}<\infty$.
	A real valued random vector is called subgaussian, if all of its one dimensional marginals are subgaussian.
\end{defi}
An obvious example of subgaussian random variables are Gaussian random variables, as the expectation in Definition~\ref{defi:subgauss} exists for all $t<1$. Also, all bounded random variables are subgaussian.

Here, we rely on results given by Tropp in~\cite{tropp2014convex} using the results of Mendelson~\cite{koltchinskii2015bounding, mendelson2014learning}.
We will consider problems of the form
\begin{equation}\label{eq:minNoise}
\min \Vert z\Vert_{TV}\,\text{ such that } \Vert Az - y\Vert \leq\varepsilon,
\end{equation}
where $A$ is supposed to be a matrix with independent subgaussian rows. Furthermore, we denote the exact solution by $x_0$, i.e., $Ax_0 = y$.
We pose the following assumptions on the distribution of the rows of $A$. 
\begin{enumerate}
	\item[(M1)] $\Ex A_i = 0$,
	\item[(M2)] There exists $\alpha>0$, such that for all $u\in\S^{n-1}$ it holds that $\Ex|\langle A_i,u\rangle|\geq\alpha$,
	\item[(M3)] There is a $\sigma>0$, such that for all $u\in\S^{n-1}$ it holds that $\Prob(|\langle A_i,u\rangle|\geq t)\leq2\exp(-t^2/(2\sigma^2))$,
	\item[(M4)] The constant $\rho:= \frac\sigma\alpha$ is small.
\end{enumerate}

Then the small ball methods yields the following recovery guarantee (we present the version of \cite{tropp2014convex}).
{\theo\label{thm:TroppRecovery} Let $x_0\in\R^n$ and $A\in\R^{m\times n}$ be a subgaussian matrix satisfying (M1)-(M4) above. Furthermore, set $y = Ax_0+e$, where $\Vert e\Vert\leq \varepsilon$ denotes the (bounded) error of the measurement.
	Then the solution $\hat x$ of~\eqref{eq:minNoise} satisfies 
	\[\Vert \hat x - x_0\Vert \leq \frac{2\varepsilon}{\max\{c\alpha\rho^{-2}\sqrt{m} - C\sigma w(S(K,x_0))-\alpha t,0\}}\]
	with probability exceeding $1-\ee^{-ct^2}$. $D(K,x_0)$ denotes the descent cone of the set $K$ at $x_0$, as defined in the previous section.}

From this we see that, provided \[m\geq \tilde C\rho^6 w^2(S(K,x_0)),\] we obtain stable reconstruction of our original vector from~\eqref{eq:minNoise}. 
Note that the theorem is only meaningful for $t=\Ocal(\sqrt{m})$, as otherwise the denominator vanishes.

In the previous section, we have shown the inclusion $S(K,x_0)\subset \Scal$ for $x_0\in K_s^0$ and hence we have that 
\[w(S(K,x_0)\leq w(\Scal)\leq c \sqrt[4]{ns}\sqrt{\log(2n)}.\]

So we see that for $m\geq \tilde C\rho^6 \sqrt{ns}\log(2n)$, we obtain the bound

\begin{align*}\Vert \hat x - x_0\Vert &\leq \frac{2\varepsilon}{\max\{c\alpha\rho^{-2}\sqrt{\tilde C} \rho^3  \sqrt[4]{ns}\sqrt{\log(2n)} - C\sigma \sqrt[4]{ns}\sqrt{\log(2n)}-\alpha t,0\}}\\
& = \frac{2\varepsilon}{\max\{\sigma(c\sqrt{\tilde C} - C) \sqrt[4]{ns}\sqrt{\log(2n)}-\alpha t,0\}}
\end{align*} 

with high probability. We conclude that, given the absolute constants $c,C$, we need to set $\tilde C\geq \tfrac{C^2}{c^2}$ in order to obtain a meaningful result. Combining all our previous discussions with Theorem~\ref{thm:TroppRecovery}, we get

\begin{theo}\label{theo:SubgaussGuarantee}
	Let $x_0\in\R^n$, $m\geq \tilde C\rho^6 \sqrt{ns}\log(2n)$ and $A\in\R^{m\times n}$ be a subgaussian matrix satisfying (M1)-(M4). Furthermore, set $y = Ax_0+e$, where $\Vert e\Vert\leq \varepsilon$ denotes the (bounded) error of the measurement, constants $c,C,\tilde C>0$ as above and $t\leq \frac{\sigma(c\sqrt{\tilde C} - C) \sqrt[4]{ns}\sqrt{\log(2n)}}{\alpha}$.
	Then the solution $\hat x$ of~\eqref{eq:minNoise} satisfies 
	\[\Prob\left(\|\hat x - x_0\| > \frac{2\eps}{\sigma(c\sqrt{\tilde C} - C) \sqrt[4]{ns}\sqrt{\log(2n)}-\alpha t}\right)\leq \ee^{-ct^2}. \]
\end{theo}

We can for example set $t = \rho (c\sqrt{\tilde C} -C) \sqrt[4]{ns}$ (for $n\geq 2$) to obtain the bound

\[\Prob\left(\|\hat x - x_0\| > \frac{2\eps}{\sigma(c\sqrt{\tilde C} - C) \sqrt[4]{ns}(\sqrt{\log(2n)}-1)}\right)\leq \ee^{-\tilde c \rho \sqrt{ns}}.\]

For example for i.i.d. standard Gaussian measurements, the constant $\rho = \sqrt{\frac2\pi}$.

\begin{figure}
	\centering
	\includegraphics[width= 0.45\textwidth]{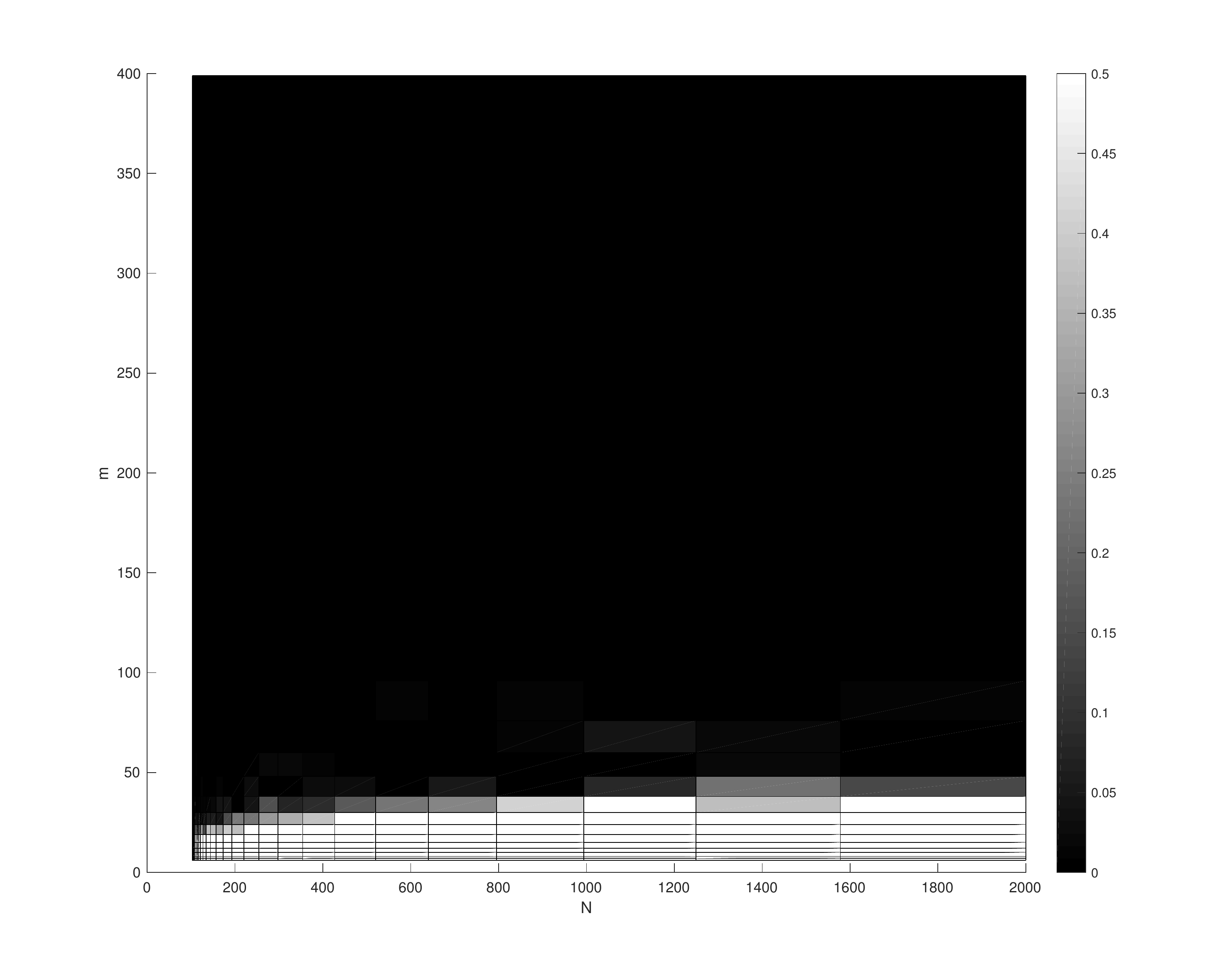}\quad
	\includegraphics[width= 0.47\textwidth]{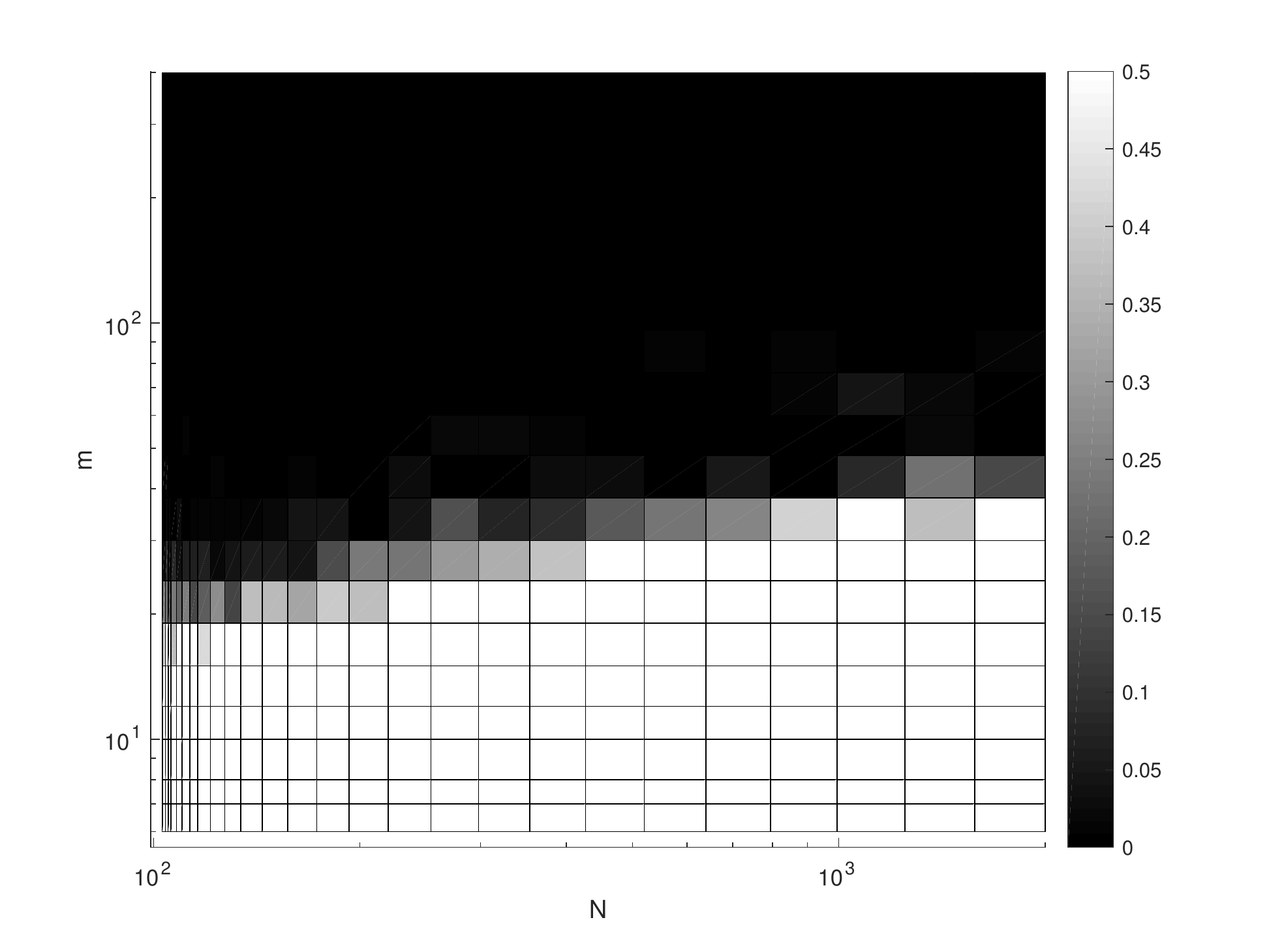}
	\caption{Average error of recovery from Rademacher measurements in $1d$ with $m$ measurements and ambient dimension $N$ for fixed cosparsity level $s = 5$. Left: linear axis scaling, Right: logarithmic axis scaling. The slope of the phase transition in the log-log plot is observed to be about $\frac14$.}
	\label{fig:Phase}
\end{figure}

Note that in the case of noisefree measurements $\eps = 0$, Theorem~\ref{theo:SubgaussGuarantee} gives an exact recovery result for a wider class of measurement ensembles with high probability. Furthermore with a detailed computation of $w(S(K,x_0))$ one may be able to improve the number of measurements for nonuniform recovery. It also remains open, whether the lower bounds of Cai and Xu for the case of Gaussian measurements can be generalized to the subgaussian case. In fact, our numerical experiments summarized in Figure~\ref{fig:Phase} suggest a better scaling in the ambient dimension, around $N^{1/4}$, in the average case. We consider it an interesting problem for future work to explore whether this is due to a difference between Rademacher and Gaussian matrix entries, between uniform and nonuniform recovery, or between the average and the worst case. Also, it is not clear whether the scaling is in fact $N^{1/4}$ or if the observed slope is just a linearization of, say, a logarithmic dependence.

\section{Discussion and open problems}
As the considerations in the previous sections illustrate, the mathematical properties of total variation minimization differ significantly from algorithms based on synthesis sparsity, especially in one dimension. For this reason, there are a number of questions that have been answered for synthesis sparsity, but which are still open for the framework of total variation minimization. For example, the analysis provided in \cite{RRT12, KMR14} for deterministically subsampled partial random circulant matrices, as they are used to model measurement setups appearing in remote sensing or coded aperture imaging, could not be generalized to total variation minimization. The difficulty in this setup is that the randomness is encoded by the convolution filter, so it is not clear what the analogy of variable density sampling would be. 

Another case of practical interest is that of sparse $0/1$ measurement matrices. Recently it has been suggested that such meausurements increase efficiency in photoacoustic tomography, while at the same time, the signals to be recovered (after a suitable temporal transform) are approximately gradient sparse. This suggests the use of total variation minimization for recovery, and indeed empirically, this approaches yields good recovery results \cite{SKBBH15}. 
Theoretical guarantees, however, (as they are known for synthesis sparse signals via an expander graph construction \cite{BGIKS08}) are not available to date for this setup.

\section*{Acknowledgements}

FK and MS  acknowledge support by the Hausdorff Institute for
Mathematics (HIM), where part of this work was completed in the context of the HIM trimester program ”Mathematics of Signal Processing”, FK and CK acknowledge support by the German Science Foundation in the context
of the Emmy Noether Junior Research Group “Randomized Sensing and Quantization of Signals and Images” (KR 4512/1-1) and by the German Ministry of Research and Education in the context of the joint research initiative ZeMat. MS has been supported by the Austrian Science Fund (FWF) under Grant no.~Y760 and the DFG SFB/TRR 109 "Discretization in Geometry and Dynamics".

\bibliographystyle{alpha}
\bibliography{thebibliography}

\end{document}